# Estimating axon radius using diffusion-relaxation MRI: calibrating a surface-based relaxation model with histology

Muhamed Barakovic[1,2,3,4,5], Marco Pizzolato[6], Chantal M.W. Tax[7,3], Umesh Rudrapatna[3], Stefano Magon[5], Tim B. Dyrby[8,6], Cristina Granziera[1,2,9], Jean-Philippe Thiran[4,10,11], Derek K. Jones[3,a], and Erick J. Canales-Rodríguez[4,a,*].

[1] Translational Imaging in Neurology (ThINk) Basel, Department of Biomedical Engineering, University Hospital Basel and University of Basel, Basel, Switzerland.

[2] Department of Neurology, University Hospital Basel.

[3] Cardiff University Brain Research Imaging Centre, Cardiff University, Cardiff, Wales, United Kingdom.

[4] Signal Processing Laboratory 5 (LTS5), Ecole Polytechnique Fédérale de Lausanne (EPFL), Lausanne, Switzerland.

[5] Roche Pharma Research & Early Development, Neuroscience and Rare Diseases, Roche Innovation Center, Basel, Switzerland.

[6] Department of Applied Mathematics and Computer Science, Technical University of Denmark, Kongens Lyngby, Denmark.

[7] Image Sciences Institute, University Medical Center Utrecht, The Netherlands.

[8] Danish Research Centre for Magnetic Resonance, Centre for Functional and Diagnostic Imaging and Research, Copenhagen University Hospital, Amager & Hvidovre, Hvidovre, Denmark.

[9] Research Center for Clinical Neuroimmunology and Neuroscience Basel (RC2NB), University Hospital Basel and University of Basel, Basel, Switzerland.

[10] Radiology Department, Centre Hospitalier Universitaire Vaudois and University of Lausanne, Lausanne, Switzerland.

[11] Centre d'Imagerie Biomédicale (CIBM), EPFL, Lausanne, Switzerland.

[a] Co-last authors.

* Corresponding author: erick.canalesrodriguez@epfl.ch

**Abstract**

Axon radius is a potential biomarker for brain diseases and a crucial tissue microstructure parameter that determines the speed of action potentials. Diffusion MRI (dMRI) allows non-invasive estimation of axon radius, but accurately estimating the radius of axons in the human brain is challenging. Most axons in the brain have a radius below one micrometre, which falls below the sensitivity limit of dMRI signals even when using the most advanced human MRI scanners. Therefore, new MRI methods that are sensitive to small axon radii are needed. In this proof-of-concept investigation, we examine whether a surface-based axonal relaxation process could mediate a relationship between intra-axonal $T_2$ and $T_1$ times and inner axon radius, as measured using postmortem histology. A unique in vivo human diffusion-$T_1$-$T_2$ relaxation dataset was acquired on a 3T MRI scanner with ultra-strong diffusion gradients, using a strong diffusion-weighting (i.e., $b$=6000 s/mm$^2$) and multiple inversion and echo times. A second reduced diffusion-$T_2$ dataset was collected at various echo times to evaluate the model further. The intra-axonal relaxation times were estimated by fitting a diffusion-relaxation model to the orientation-averaged spherical mean signals. Our analysis revealed that the proposed surface-based relaxation model effectively explains the relationship between the estimated relaxation times and the histological axon radius measured in various corpus callosum regions. Using these histological values, we developed a novel calibration approach to predict axon radius in other areas of the corpus callosum. Notably, the predicted radii and those determined from histological measurements were in close agreement.

**Highlights**

- Diffusion-relaxation MRI data were acquired using a high *b*-value acquisition.
- A diffusion-relaxation model to estimate the intra-axonal $T_2$ and $T_1$ was proposed.
- The histological inner axon radius modulated the estimated relaxation times.
- A surface-based relaxation model predicted the axon radius in the corpus callosum.
- The predicted axon radii agreed with the mean effective histological radius.

**Graphical abstract (1328x531 pixels)**

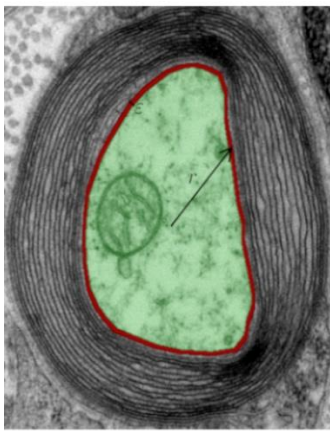
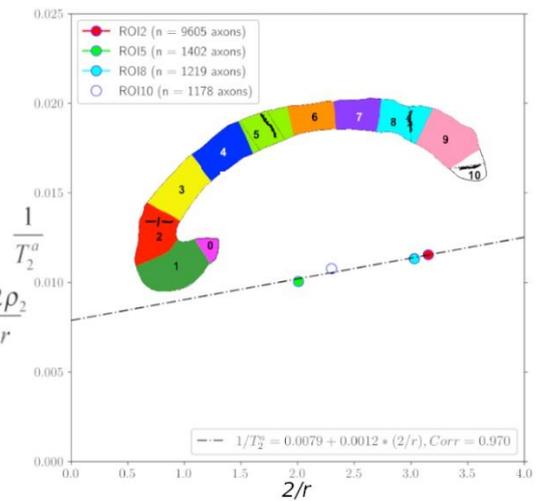

# 1. Introduction

The speed of action potentials along axons is partly determined by their radii (Goldstein and Rall, 1974). Axon radius explains the biggest variance in conduction speed, as demonstrated by previous studies (Hursh, 1939), with larger axons conducting faster than those with smaller radii (Waxman and Bennett, 1972; Costa et al., 2018; Drakesmith et al., 2019). Therefore, accurately measuring axon radii in vivo is essential for better understanding the neural mechanisms underlying brain function and their impact on diseases.

The diffusion Magnetic Resonance Imaging (dMRI) signal is sensitive to axon radii if strong diffusion encoding gradients (i.e., up to 300 mT/m in Connectom scanners (Jones et al., 2018) and 1500 mT/m in animal preclinical scanners) are used (Assaf et al., 2004, 2008; Assaf and Basser, 2005; Alexander, 2008; Dyrby et al., 2013; Duval et al., 2015; De Santis et al., 2016; Veraart et al., 2020; Barakovic et al., 2021a). However, the main limitation of this approach is that the dMRI signals from axons with radii smaller than ~1-2 μm are practically indistinguishable from each other, even when the most advanced human Connectom scanners with ultra-strong (300 mT/m) gradients are employed in the data acquisition (Nilsson et al., 2017). Today, the challenge is that the peak of the axon radius distribution per voxel is below one micrometre in most brain regions, as observed in histology. Hence, most axon radii are below the lower bound for detection (Edgar and Griffiths, 2014; Dyrby et al., 2018). For an overview of the different strategies that have been employed to measure axon radius with dMRI, the reader is referred to (Assaf and Basser, 2005; Assaf et al., 2008, 2013; Alexander et al., 2010, 2019; Dyrby et al., 2013, 2018; Novikov et al., 2019; Veraart et al., 2020; Fan et al., 2020; Jelescu et al., 2020; Barakovic et al., 2021a; Pizzolato et al., 2023).

Theoretical reasons explain the lower sensitivity of dMRI to the inner radius of smaller axons. The commonly employed model (i.e., Gaussian phase approximation in the long-pulse limit (van Gelderen et al., 1994)) predicts an intra-axonal dMRI signal attenuation that depends on the fourth power of the radius $r$. Moreover, since the measured intra-axonal signal per voxel is the sum of all the individual intra-axonal signals weighted by each axon's contribution to the signal (scaling by an extra-factor $r^2$), larger axons contribute more than smaller axons to the measured signal. After considering these two factors together, an approximate expression for the mean 'effective' dMRI-based radius $r_{eff}$ per voxel can be derived, which depends on the higher-order moments of the unknown axon radius distribution. The resulting analytical expression $r_{eff} \approx \left( \langle r^6 \rangle / \langle r^2 \rangle \right)^{\frac{1}{4}}$ (where $\langle \ \rangle$ denotes the average over the distribution) demonstrates that the estimate is heavily weighted by the right-hand tail of the axon radius distribution (Burcaw et al., 2015; Veraart et al., 2020).

Consequently, the estimated mean axon radius is mainly affected by the bigger axons from the fractions of axons larger than the lower bound. This explains why estimations may appear overestimated compared to histology (Alexander et al., 2010; Dyrby et al., 2018).

Finding another source of MRI contrast sensitive to the size of axons smaller than the diffusion resolution limit is essential. Various studies in porous media have demonstrated that the interaction between the water molecules and the confining pore surface reduces the observed transverse $T_2$ relaxation time (Brownstein and Tarr, 1977). This surface-based relaxation mechanism allows pore size to be estimated (Hurlimann et al., 1994; Slijkerman and Hofman, 1998; Sørland et al., 2007; Mohnke and Hughes, 2014; Müller-Petke et al., 2015). Notably, a similar $T_2$ relaxation model to predict the size of cells was proposed previously (Brownstein and Tarr, 1979), and the idea of applying it to estimate the axon radius was suggested by (Kaden and Alexander, 2013). However, there is a lack of validation studies to demonstrate whether the inner axon radius modulates the intra-axonal relaxation times. This might be explained by the fact that approaches to estimating the intra-axonal relaxation times have only been developed recently (Veraart et al., 2018; McKinnon and Jensen, 2019; Barakovic et al., 2021b; Tax et al., 2021; Pizzolato et al., 2022). Furthermore, to our knowledge, no dataset is available that offers the combined histological information and relaxometry MRI data from the same sample, which are necessary for the estimation and comparison of these parameters.

The dMRI signals arising from the intra-axonal space can be isolated if a sufficiently high b-value is employed (i.e., $b$>4000 s/mm$^2$ for in vivo data), which significantly attenuates the signal from spins experiencing large displacements (Jensen et al., 2016; McKinnon and Jensen, 2019). As the confining axonal geometry restricts the self-diffusion motion of spins inside axons (assuming a slow exchange between the intra- and extra-axonal spaces), the strongly diffusion-weighted MRI signal should come from the intra-axonal spins. Thus, it is possible to fit a diffusion-relaxation model of intra-axonal relaxation to strongly diffusion-weighted MRI data collected at multiple diffusion gradient directions and different echo times. This approach, combined with taking the spherical mean (orientational average), was employed previously to estimate the mean intra-axonal $T_2$ time per voxel (McKinnon and Jensen, 2019) and bundle (Barakovic et al., 2021b), unconfounded by fibre orientation effects.

This proof of concept study investigates whether the intra-axonal $T_2$ and $T_1$ relaxation times are related to the inner axon radius and whether they can be employed to predict the mean effective radius. To do this, (1) we implemented two acquisition protocols and measured diffusion-$T_1$-$T_2$ and diffusion-$T_2$ weighted MRI data from three healthy volunteers, one of them scanned using both sequences; (2) we employed a diffusion-relaxation model to enable the estimation of both intra-axonal $T_2$ and $T_1$ relaxation times by using the spherical mean signals from the acquired data; (3) we fitted the estimated relaxation times to a surface-based relaxation model that depends on the histological axon radius; (4) using histology from some brain regions we calibrated the surface-based relaxation model to enable predicting axon radius in other brain regions, and (5) we compared the MRI-based estimated axonal radii with those obtained from two postmortem histological human brain datasets in several regions in the midsagittal Corpus Callosum (CC) cross-section. Additional details are provided at the end of the next section.

## 2. Theory

**2.1 Surface-based relaxation model**

Inspired by the standard surface-based relaxation model used in porous media (Zimmerman and Brittin, 1957; Brownstein and Tarr, 1979), we propose the following model described in Figure 1 and Eqs. (1)-(2). We assume that in the intra-axonal space, there are two distinct water pools in fast exchange (Zimmerman and Brittin, 1957): the surface water immediately adjacent to the axonal membrane, e.g., see (Le Bihan, 2007), and the cytoplasmic water (i.e., axoplasm). The $T_2$ and $T_1$ relaxation times of the surface water are shorter because this water layer is in a more ordered state (both spatially and orientationally) than pure water (Halle, 1999; Finney et al., 2004) and the cytoplasmic water, due to the strong water-tissue interactions (Levy and Onuchic, 2004; Zhang et al., 2007). Moreover, the relaxation times of the cytoplasmic water are expected to be smaller than those of pure water and Cerebrospinal fluid because the water molecules in this pool could interact with cytoskeletal elements and a higher number of macromolecules (Beaulieu, 2002). The fast exchange assumption is reasonable if we consider that water molecules, on average, travel distances much larger than the axon radius for typical diffusion and echo times, as employed in this study.

According to the general model provided by (Zimmerman and Brittin, 1957), the inverse of the observed intra-axonal $T_2$ can be modelled by the linear combination of the inverse relaxation times of the surface water and the cytoplasmic water pools, weighted by their volume fractions. Although the volume of the surface water layer is much smaller than the total intra-axonal volume, its $T_2$ time ($T_2^s$) is much shorter than that of the cytoplasmic water ($T_2^c$). It thus could have a non-negligible impact on the observed intra-axonal ($T_2^a$) time. These assumptions are summarised in the following model:

$$\frac{1}{T_2^a} = \frac{V - S\varepsilon}{V}\frac{1}{T_2^c} + \frac{S\varepsilon}{V}\frac{1}{T_2^s}$$
$$\approx \frac{1}{T_2^c} + \frac{2\varepsilon}{r}\frac{1}{T_2^s} \qquad (1)$$
$$= \frac{1}{T_2^c} + \frac{2\rho_2}{r},$$

where $\rho_2 = \varepsilon/T_2^s$ is the T$_2$ surface relaxivity; $S$ is the surface area of the axonal membrane; $V$ is the intra-axonal volume; $\varepsilon$ is the thickness of the water layer. Note that when assuming a cylindrical axonal geometry, as commonly done in dMRI, the surface-to-volume ratio depends on the inner axon radius, $S/V = 2/r$. An equivalent expression was obtained for the intra-axonal T$_1$ time

$$\frac{1}{T_1^a} \approx \frac{1}{T_1^c} + \frac{2\rho_1}{r}, \qquad (2)$$

where $\rho_1 = \varepsilon/T_1^s$ is the longitudinal surface relaxivity.

Insert Figure 1 (1.5 columns)

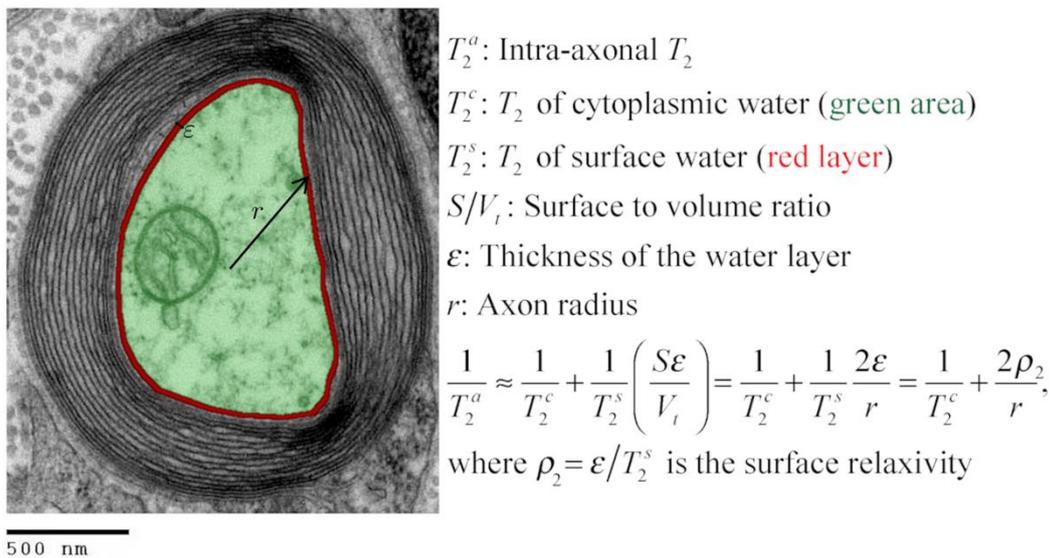

**Figure 1**. Transmission electron micrograph of a myelinated axon (adapted) illustrating the employed relaxation model for the intra-axonal space, composed of two pools (arbitrarily coloured in green and red for illustrative purposes) in fast exchange (Zimmerman and Brittin, 1957). This model is equivalent to the Brownstein-Tarr model in the fast diffusion limit (Brownstein and Tarr, 1977). The structured water (Le Bihan, 2007) adjacent to the inner axon surface (red) has a shorter T$_2$ than the cytoplasmic water (green). As the cytoplasmic water (i.e., axoplasm) interacts with large proteins, organelles, and cytoskeletal elements (LoPachin et al., 1991; Beaulieu, 2002), its T$_2$ is shorter than pure water. An equivalent model was assumed for the $T_1$ relaxation. [This transmission electron micrograph was deposited into the public domain by the Electron Microscopy Facility at Trinity College].

**2.2 Axon radius estimation from intra-axonal relaxation times**

By inverting Eqs. (1) or (2) it is possible to predict the inner axon radius from the estimated intra-axonal $T_2^a$ and $T_1^a$ relaxation times, respectively

$$r \approx \frac{2\rho_2}{\frac{1}{T_2^a} - \frac{1}{T_2^c}},$$

$$r \approx \frac{2\rho_1}{\frac{1}{T_1^a} - \frac{1}{T_1^c}}. \tag{3}$$

However, this approach requires knowing $T_2^c$ and $\rho_2$ or $T_1^c$ and $\rho_1$ in advance. As these parameters are unknown and cannot be estimated for each brain voxel without additional data, here we propose a histologically-informed calibration approach to calculate them.

The calibration is based on assuming that any dependence of $T_2^c$ or $T_1^c$ on the axon radius, owing to potential changes in the intra-axonal structure with the axon size (e.g., density of proteins, organelles, and cytoskeletal elements), is weak and can be neglected. That is, the dependence of $T_2^a$ and $T_1^a$ on the axon radius is dominated by the surface-to-volume ratio terms in Eqs. (1) and (2). Therefore, we assume that $T_2^c$, $\rho_2$, $T_1^c$ and $\rho_1$ are constant across axons with different sizes. Nevertheless, we noted that the calibration process is equally helpful in predicting axon radius when $T_2^c$ or $T_1^c$ linearly varies with the radius. In that case, the linear models (Eqs. (1) and (2)) can be rewritten in terms of two alternative parameters. For more details, see the discussion subsection "Is the cytoplasmic $T_2$ constant?".

In this study, we collected in vivo diffusion-$T_1$-$T_2$ MRI data in a human brain to estimate $T_2^a$ and $T_1^a$. We employed a reduced diffusion-$T_2$ relaxation sequence to validate our model further by scanning the same subject and two additional healthy volunteers, which allowed us to estimate $T_2^a$. Subsequently, we used histological information from four regions of interest (ROIs) located in the CC of a postmortem human brain to measure the mean histological axon radii. The mean intra-axonal relaxation times and histological axonal radii estimated in the four ROIs were combined to estimate $T_2^c$ and $\rho_2$, and $T_1^c$ and $\rho_1$ via linear regression (i.e., calibration step) from Eqs. (1) and (2). Then, using the calibrated parameters, we predicted axon radius in another eleven CC ROIs for each scanned subject via Eq. (3). Finally, we employed a second histological dataset containing data from nine postmortem human brains to further validate our results. All the details are provided in the Methods section.

## 3. Methods

### 3.1 Intra-axonal diffusion-relaxation models

As in (McKinnon and Jensen, 2019), we assume that for $b = 6000$ s/mm$^2$ the in vivo dMRI signal comes from the intra-axonal space. Thus, the diffusion-$T_1$-$T_2$ relaxation model for the measured signal $M$ for a given $b$, diffusion gradient unit vector $\hat{g}$, echo time (*TE*), repetition time (*TR*), and inversion time (*TI*) is

$$M(b, \hat{g}, TE, TI) = kPDf_a M_a(b, \hat{g}) \exp\left(-\frac{TE}{T_2^a}\right) \left|1 - 2\exp\left(-\frac{TI}{T_1^a}\right) + \exp\left(-\frac{TR}{T_1^a}\right)\right| + \eta, \quad (4)$$

where $k$ is a scalar that depends on the MRI machine, pulse sequence, image-reconstruction algorithm, digital converter, etc.; *PD* is the proton density; $f_a$ is the intra-axonal water volume fraction; $M_a(b, \hat{g})$ denotes the orientation-dependent diffusion-weighted signal from the intra-axonal compartment; $\eta$ is the experimental noise, assumed to be additive; $|x|$ denotes the absolute value of $x$; $T_2^a$ and $T_1^a$ are the intra-axonal relaxation times.

Following the approach of (Edén, 2003; Lasič et al., 2014; Kaden et al., 2016b, 2016a), Eq. (4) can be simplified by computing the orientation-averaged spherical mean signal $\bar{M}$ as:

$$\bar{M}(b, TE, TI) \approx K \exp\left(-\frac{TE}{T_2^a}\right) \left|1 - 2\exp\left(-\frac{TI}{T_1^a}\right) + \exp\left(-\frac{TR}{T_1^a}\right)\right|, \quad (5)$$

where $T_2^a$ and $T_1^a$ are the parameters to be estimated, along with the constant $K$ (per voxel) that is proportional to the intra-axonal water volume fraction (i.e., $K = kPDf_a \bar{M}_a(b, \hat{g})$); it also depends on the intra-axonal diffusivities via $\bar{M}_a$.

It is important to note that in Eqs. (4)-(5), the T$_1$ relaxation terms follow the standard relaxation model (Bydder et al., 1998), which assumes an ideal inversion pulse (Pykett et al., 1983; Barral et al., 2010). Other acquisition sequences may require different models. For a comprehensive review of alternative relaxometry sequences and models, please refer to (Stikov et al., 2015).

The diffusion-relaxation model in Eq. (5) is a more general version of the model proposed by (McKinnon and Jensen, 2019) for an inversion recovery sequence incorporating T$_1$ relaxation. The diffusion-T$_2$ model for dMRI data collected at multiple *TE*s (McKinnon and Jensen, 2019) without considering T$_1$ effects is

$$\bar{M}(b, TE, TI) \approx K \exp\left(-\frac{TE}{T_2^a}\right). \tag{6}$$

**3.2 MRI data acquisition and preprocessing**

Human brain MRI data were acquired from three healthy volunteers, and one of them was scanned twice on a Siemens Connectom 3T system with 300 mT/m diffusion gradients (Cardiff University Brain Research Centre, Wales, UK). The ethics committee approved the study, and the participant provided written informed consent.

Two diffusion-relaxation protocols were implemented. A longer diffusion-T$_1$-T$_2$-weighted imaging sequence was designed to obtain independent estimates of the axon radius from the first subject's intra-axonal T$_1$ and T$_2$ times (male, 28 years old). A reduced diffusion-T$_2$ protocol was employed to scan three subjects (age-range=28-39 years, mean-age=32.3±4.8 years, males), including the first subject that was also scanned with the longer sequence. Accordingly, for the second sequence, the axon radii were estimated from the intra-axonal T$_2$ times.

The diffusion-T$_1$-T$_2$ relaxation sequence comprised four images with *b*=0 s/mm$^2$ and 48 diffusion directions at *b*=6000 s/mm$^2$ (diffusion gradient, 275 mT/m; diffusion times Δ/ δ=22/8 ms) for each of the following nine (*TE*, *TI*) combinations (in ms): (80, 200), (110, 200), (110, 331), (150, 200), (80, 906), (110, 906), (110, 1500), (150, 906), (150, 1500). The *TI*s were chosen empirically from relatively small to large values to obtain maps with different visual contrasts without nullifying the WM signal. The lowest *TE* was set to minimise the contribution of the myelin water (Mackay et al., 1994) to the measured signal, and the largest *TE* was chosen as a trade-off between image contrast and noise. For each (*TE*, *TI*) pair, one additional image with *b* = 0 s/mm$^2$ and opposite phase encoding direction was acquired to correct susceptibility distortions (Andersson et al., 2003; Andersson and Sotiropoulos, 2016). Figure 2 shows the nine pairs of *TE*s and *TI*s. The *TR* was 5000ms, and the voxel size was 2.5x2.5x3.5 mm$^3$. Ten slices were acquired with matrix size and field of view of 88x88 and 220x220 mm$^2$, respectively. The acceleration factor was 2, and the total acquisition time was 42 minutes.

The diffusion-$T_2$ protocol employed a dMRI sequence that was repeated by changing the *TE*, using the following four values *TE*s=(73, 93, 118, and 150) ms with *TR*=4100ms. The other sequence parameters (i.e., acceleration factor, diffusion times, *b*-value, diffusion directions, number of *b*0s images, diffusion gradient strength, matrix size, and field of view) were equal to those employed in the previous diffusion-$T_1$-$T_2$ sequence. The number of slices was 46, and the voxel size was 2.5x2.5x2.5 mm$^3$. The acquisition time per *TE* was 5 minutes, and the total scan time was 20 minutes.

Additionally, a structural $T_1$–weighted (T1w) image was collected for each subject using a 3D MPRAGE sequence with the following parameters: *TR* = 2300 ms, *TE* = 2 ms, *TI*=857 ms, voxel size = 1 mm isotropic, and flip angle=9°, for the purposes of spatial normalisation.

The nine diffusion-$T_1$-$T_2$ 4D volumes with different *TE*s and *TI*s, and the four diffusion-$T_2$ 4D volumes with different *TE*s were preprocessed separately in the following order: (1) noise level estimation and removal using the MP-PCA method (Veraart et al., 2016) by using the matrix centring and patch-based aggregation options (Manjon et al., 2013), as implemented in dipy (Garyfallidis et al., 2014) (https://dipy.org/); (2) attenuation of the Rician-noise dependent bias in the signal by implementing the postprocessing correction scheme proposed by (Gudbjartsson and Patz, 1995) and (3) motion, geometric distortions, and eddy current corrections using the 'topup' and 'eddy' tools included in FSL (Andersson et al., 2003; Andersson and Sotiropoulos, 2016).

Insert Figure 2 (1 column)

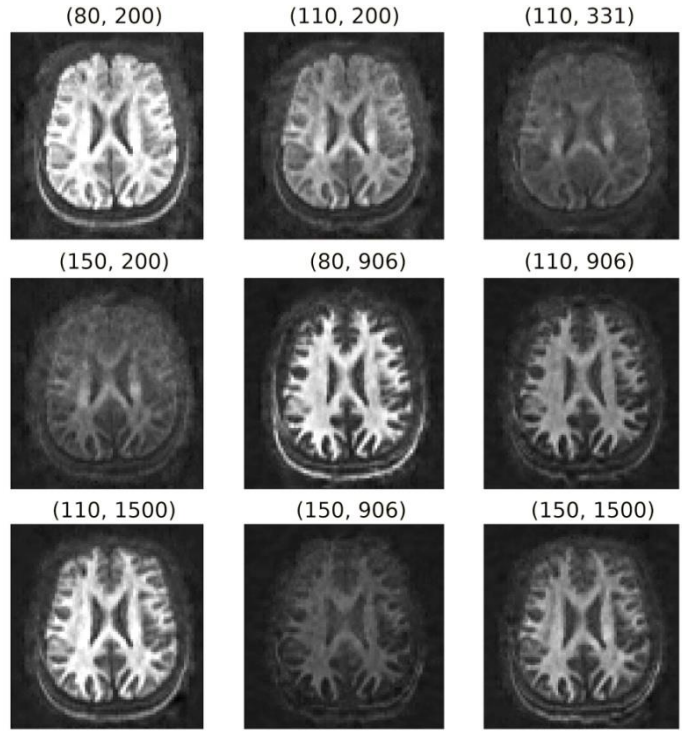

**Figure 2**. Orientation-averaged spherical mean signals for each pair of *TE* and *TI* (TE, TI) in ms. These images were used to fit the diffusion-relaxation model in Eq. (5).

### 3.3 Estimation of the intra-axonal relaxation times

Diffusion-$T_1$-$T_2$ model: after computing the spherical mean signal for each pair of the preprocessed diffusion-$T_1$-$T_2$ datasets with different *TE*s and *TI*s (see Figure 2), the intra-axonal relaxation times were estimated by fitting the diffusion-relaxation model in Eq.(5) using the 'L-BFGS-B' method for bound constrained minimisation included in the Scipy python library (Virtanen et al., 2020) (https://docs.scipy.org), with the following bounds: $0 \leq K < \infty$, $40 \leq T_2^a(ms) \leq 2000$, and $300 \leq T_1^a(ms) \leq 5000$. The bounds for the intra-axonal relaxation times were chosen to be higher and lower than those expected for the myelin water and Cerebrospinal fluid (Mackay et al., 1994; Labadie et al., 2014), respectively.

Diffusion-$T_2$ model: the estimation was performed by fitting the diffusion-relaxation model in Eq. (6) to the spherical mean signals estimated from the diffusion-$T_2$ data, using the L-BFGS-B method (Virtanen et al., 2020) with the following bounds: $0 \leq K < \infty$, $40 \leq T_2^a(ms) \leq 2000$.

### 3.4 Histological samples

Two histological datasets were employed. The first one contains two histological samples measured on the same subject. The first sample, which we call 'Histology1', was measured and reported by (Caminiti et al., 2009). For completeness, we provide a summary of the histological procedures. Axon radii were measured in four regions of interest (i.e., ROI2, ROI5, ROI8, and ROI10) in the midsagittal

CC cross-section of a postmortem human brain (female, 63 years old). These ROIs include axons connecting the prefrontal, motor, parietal and visual cortices, respectively. All analyses were performed with Neurolucida 7 software (MBF Biosciences) and a digital camera-mounted Olympus BX51 microscope. Three sagittal blocks of the CC were removed from the brain. The sample was immersion-fixed in 4% (w/vol) paraformaldehyde in phosphate-buffered saline solution within 27–30h of death, cryoprotected, cut frozen, and stained for myelin. Axons were sampled within 112×87 µm$^2$ frames divided into 25-µm squares. The axonal profiles were chosen for measurement if they presented a dark complete or nearly complete myelin ring with a clear centre. Longitudinally cut axons were excluded, and the radius of slightly obliquely cut axons (which appeared as ellipses) was approximated to its smallest radius. Since fixation artefacts were frequent, the sampling was restricted to profiles that could be followed through the thickness of the whole section. Limitations of the optical microscopy prevented measurement of axons radius smaller than ~0.17 µm. A different number of axons were measured per ROI, ranging from 1178 (ROI10) to 9605 (ROI2) axons. No correction for shrinkage effects was applied to the measured radii because accurate shrinkage estimates were unavailable. For more technical details, see (Caminiti et al., 2009). The second sample, which we call 'Histology2', was measured by the same team (Prof. Giorgio Innocenti) using the same material and following the same sampling procedure. The main difference was that this time, eleven ROIs (i.e., ROI0-ROI10) encompassing the whole midsagittal CC cross-section were analysed, and the number of measured axons per ROI was smaller: from 153 (ROI5) to 720 (ROI1) axons. It is important to note that the spatial locations of ROI2, ROI5, ROI8, and ROI10 are the same in both histological samples. However, the sampling procedure employed in the Histology2 sample was repeated without including the axons measured in the Histology1 sample. The anatomical location of the ROIs in both histological samples and the number of measured axons per ROI are displayed in Figure 3.

The second histological dataset, which we call 'Histology3', was reported by (Wegiel et al., 2018). This electron microscopic study of the CC included nine control subjects (age-range=4-52 years old; mean-age=26.3 ± 15.8 years; postmortem-interval=15 ± 6.6 h; six males and three females) with well-preserved CC ultrastructure. Each brain was fixed in 10% buffered formalin for at least three months, washed for 24 h in water to remove fixer, dehydrated, embedded in celloidin, and cut into 200-µm-thick sections. Samples were oriented to cut axons perpendicularly to the long axon axis and stained with a 2% solution of p-phenylenediamine. Each section was stained with uranyl acetate and photographed at a magnification of 15,000x using a Hitachi H7500 transmission electron microscope with an Advanced Microscopy Technique (AMT) Image Capture Engine (Danvers, MA). Axons from five different segments (i.e., I, II, III, IV, and V) of the midsagittal CC cross-sections of the nine control subjects were measured. The study was limited to myelinated axons, which were better preserved than non-myelinated axons. For each case, 12 electron micrographs were used, and background correction was applied to reduce the risk of distortions during image analysis. Axons were manually delineated, and the Image J software was employed to obtain the axon radius (Feret's radius, μm) and area (μm$^2$). No correction for shrinkage effects to the measured radii was reported. The total number of axons measured in the nine control subjects was 15,085 (1676 per subject, and 335 per segment, on average). For additional details, see (Wegiel et al., 2018).

We note that the CC segments employed in both histological datasets (i.e., Histology1-Histology2 and Histology3) are related. Segment I (Histology3) approximately corresponds to the union of ROI0, ROI1, and ROI2 (Histology1-Histology2); segment II is located around ROI3 and ROI4; the union of segments III and IV is similar to the union of ROI5, ROI6, and RO7; and segment V covers ROI8, ROI9 and ROI10. These relationships were used to compare the histological estimates from both studies and the MRI-based radius estimates.

**3.5 Estimation of the mean histological effective radius**

For each ROI of each sample, we estimated the mean histological axon radius. However, as the mean axon radius estimated from MRI generally differs from the mean histological radius (Burcaw et al., 2015; Veraart et al., 2020), we derived an approximate expression for the mean effective radius for our diffusion-relaxation models, finding that $r_{eff} \approx \langle r^2 \rangle / \langle r \rangle$, which differs from the previous result reported in (Burcaw et al., 2015; Veraart et al., 2020). (The complete derivation is reported in the Appendix section.) This key result shows that the mean effective radius derived from our model is not heavily weighted by the tail of the axon radius distribution as that in (Burcaw et al., 2015; Veraart et al., 2020). Consequently, we used this expression to estimate the mean effective histological axon radius for each CC ROI in both samples (see Figure 3), which was compared with the MRI-predicted mean radius.

In order to estimate the effective radius, knowing both the mean histological axon radius and the mean squared radius is required. For the Histology1 and Hostology2 samples, these values were calculated from the whole radius distribution per ROI. We don't have access to the radius distributions of the Histology3 sample. Fortunately, in that study, the mean histological radius and the mean axon area were reported (Wegiel et al., 2018). We used the mean axon area to estimate the mean squared radius assuming a circular geometry.

Insert Figure 3 (1 column)

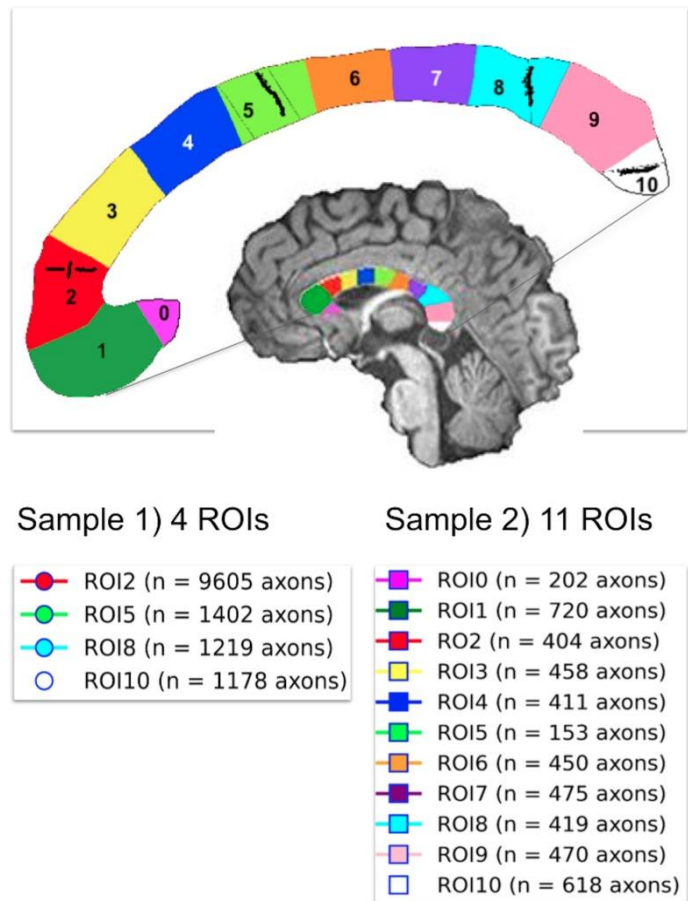

**Figure 3**. Anatomical location of the two independent histological samples of the first histological dataset (Histology1 and Histology2) taken from eleven regions of interest (ROIs) in the Corpus Callosum. The number of studied axons per sample and ROI are reported for each case. The second sample (Histology2) consisted of axons not included in the first sample (Histology1).

**3.6 Spatial registration**

The histologist that measured the axons in the Histology2 sample drew the locations of the eleven histological ROIs on the structural T1w image of the subject scanned using the diffusion-$T_1$-$T_2$ sequence, which we used to create a cluster mask. Therefore, we used that T1w image as a reference to spatially register the estimated parameter maps for all the subjects (i.e., intra-axonal relaxation times and $K$ maps). The same affine registration matrix and nonlinear deformation field were applied to each subject's estimated parameter map. These registration parameters were determined by nonlinearly registering the estimated $K$ map (whose visual appearance is similar to a T1w image, e.g., see Figure 4 in the results section) to the reference T1w image. The registration was carried out using the state-of-the-art (Klein et al., 2009) Symmetric Normalization (SyN) method (Avants et al., 2008) implemented in the ANTs software (*ANTsPy*: https://github.com/ANTsX/ANTsPy). Before the registration, we corrected the $K$ map and T1w image for spatial intensity variations due to B1-Radiofrequency field inhomogeneities using FSL (Smith et al., 2004). All the registered images were visually inspected to verify the accuracy of the normalisation procedure. All the subsequent analyses employed the registered maps. Furthermore, the ROIs were eroded to remove peripheral voxels that do not correspond to the corpus callosum and are affected by partial volume effects with surrounding tissue and CSF.

The number of voxels included in each ROI ranges from 170 (ROI0) to 604 (ROI1) in the cluster mask defined in the reference T1w image. The equivalent number of voxels in the native space of the diffusion-$T_1$-$T_2$ MRI data with a lower spatial resolution (obtained after applying the resulting nonlinear inverse registration to the cluster mask) ranges from 10 (ROI10) to 20 (ROI1).

**3.7 Calibration step to predict axon radii**

The first sample of the first histological dataset (Histology1) was employed to estimate the unknown parameters of the surface-based relaxation models in Eqs. (1)-(2). These equations were fitted independently using the mean intra-axonal $T_2$ and $T_1$ times and the mean effective histological radii estimated in the same four ROIs of the Histology1 sample. The fitting allowed us to determine the cytoplasmic $T_2^c$ and $T_1^c$ times and the surface relaxivity coefficients $\rho_2$ and $\rho_1$, which best explain the data in these regions. This was done by fitting the linear equation $y = mx + n$, where $y = 1/T_2^c$ and $x = 2/r$ for values from the four CC ROIs. Note that these parameters can be estimated as $\rho_2 = m$ and $T_2^c = 1/n$. A similar independent linear model was used to fit the $T_1$ data for estimating $\rho_1$ and $T_1^c$.

Subsequently, we predicted the mean effective axon radii, using Eq. (3), in the eleven CC ROIs of the second sample of the first dataset (Histology2) and the CC segments defined in the second histological dataset (Histology3). The forecasted and histological axon radii were compared using a linear regression model. The linear relationship among the parameters was quantified and tested by the slope and intercept of the fitted regression line and the Pearson correlation coefficient. It is important to mention that when there are two variables, such as in our study, the p-value of the slope of the regression line and the p-value of the correlation coefficient are identical. Therefore, to avoid redundancy, we have reported only the p-values of the slopes in our findings. In the Results section, we present the raw p-values without applying the correction for multiple comparisons. However, in the Discussion section, we mention the analyses that have survived the Bonferroni correction.

## 4. Results

**4.1 diffusion-$T_1$-$T_2$ and Histology1-Histology2 data**

Figure 4 shows the $T_2^a$, $T_1^a$, and $K$ maps estimated from the in vivo diffusion-$T_1$-$T_2$ MRI data for different brain slices. The estimated relaxation times are within the expected range for white matter. The values in the whole medial part of the CC were distributed in the following ranges: $70 < T_2^a (ms) < 130$, $650 < T_1^a (ms) < 760$.

Insert Figure 4 (1.5 columns)

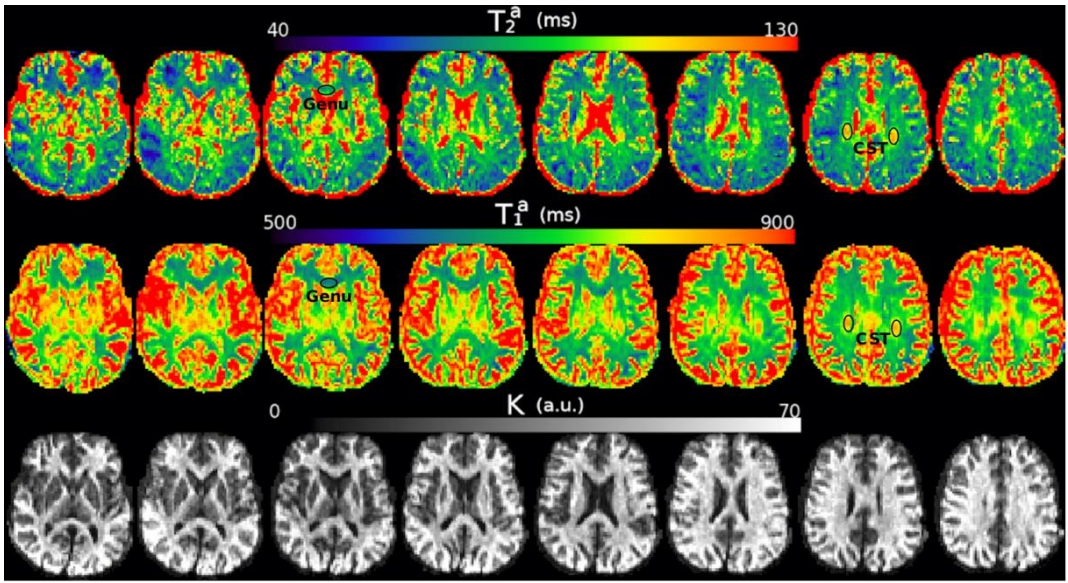

**Figure 4**. Axial slices of the $T_2^a$, $T_1^a$, and $K$ maps estimated from the in vivo diffusion-$T_1$-$T_2$ relaxation MRI data in native space (i.e., before registering the images to the reference T1w image). Note that the intra-axonal relaxation times are only meaningful in the white matter because the assumptions underlying the estimation method are invalid in grey matter or CSF. The values of $K$ (in arbitrary units) are higher in the white matter because this parameter is proportional to the intra-axonal volume. We highlight two regions with different intra-axonal relaxation times: the genu of the Corpus Callosum and the corticospinal tract (CST).

The results of the calibration step are depicted in Figure 5. It shows the regression line fitting the inverse of the mean intra-axonal $T_2$ per ROI to the inverse of the mean histological radius in the four ROIs of the Histology1 sample (for more details, see Figure 1), employing the surface-based relaxation model in Eq. (1), as described in the subsection "Calibration step to predict axon radii". The correlation coefficient between both variables was 0.97, and the p-value of the slope (i.e., for a hypothesis test whose null hypothesis is that the slope is zero) was p=0.03. We found the calibrated parameters $T_2^c \approx 126.97$ ms and $\rho_2 \approx 1.16$ nm/ms from the estimated coefficients.

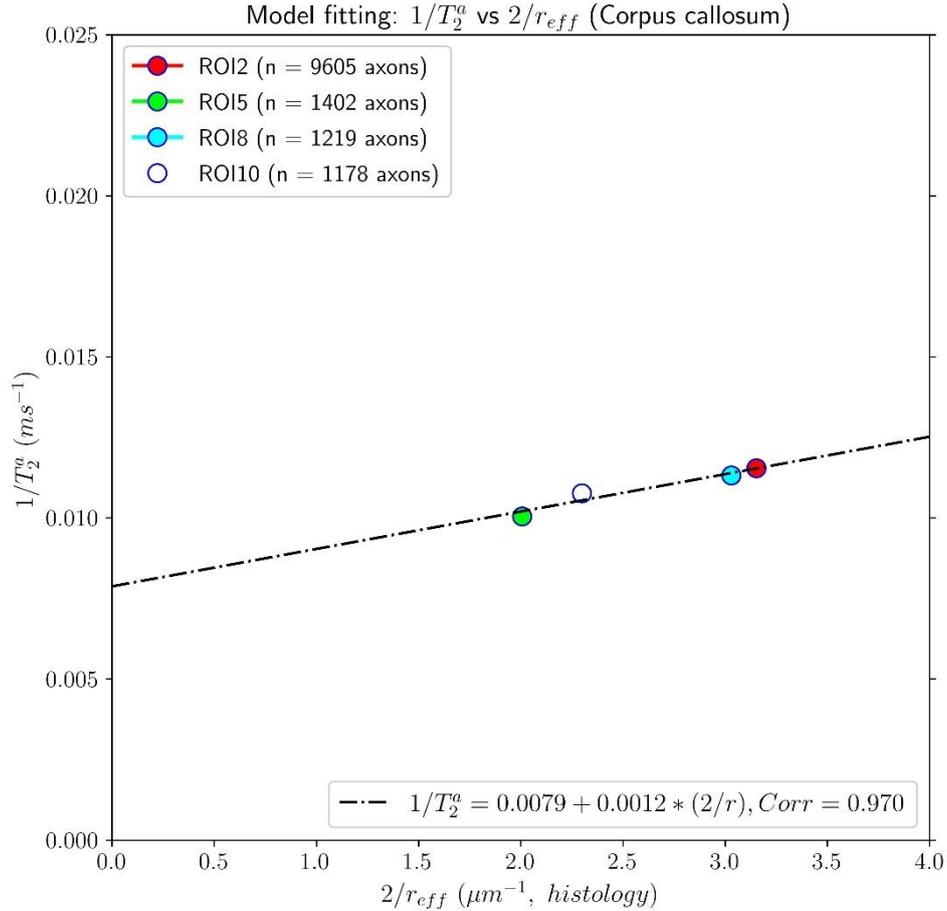

**Figure 5**. Linear fitting of the inverse of the intra-axonal T$_2$ times (y-axis) estimated from the in vivo diffusion-T$_1$-T$_2$ MRI data to the inverse of the inner axon radius (x-axis) measured from the first histological sample (Histology1) of the first histological dataset. The scatter plot depicts the mean values computed for all the voxels inside four corpus callosum (CC) regions of interest, corresponding to ROI2, ROI5, ROI8, and ROI10 in the Histology1 sample. The number of axons sampled for each CC ROI is displayed in the legend. The intercept and slope of the regression line were 0.0079 ms$^{-1}$ and 0.00116, respectively. The slope of the regression line was significantly different from zero (p=0.030).

In Figure 6, we compare the effective histological radii in the eleven ROIs of the Histology2 sample and those predicted using the intra-axonal T$_2$ times estimated from the in vivo diffusion-T$_1$-T$_2$ MRI data (Eq. (3)). The intercept and slope of the regression line were 0.026 µm and 1.055, respectively; the correlation coefficient was 0.676, and the p-value for the slope and the correlation was p=0.022. To further investigate the data, we analysed a subset of seven ROIs, excluding the four ROIs in the same locations as those in the Histology1 sample. We obtained a slightly higher intercept of 0.12 µm and a smaller slope of 0.89 compared to the analysis conducted with eleven ROIs. The resulting correlation coefficient decreased to 0.557, and the p-value for the slope was not significant, p=0.19.

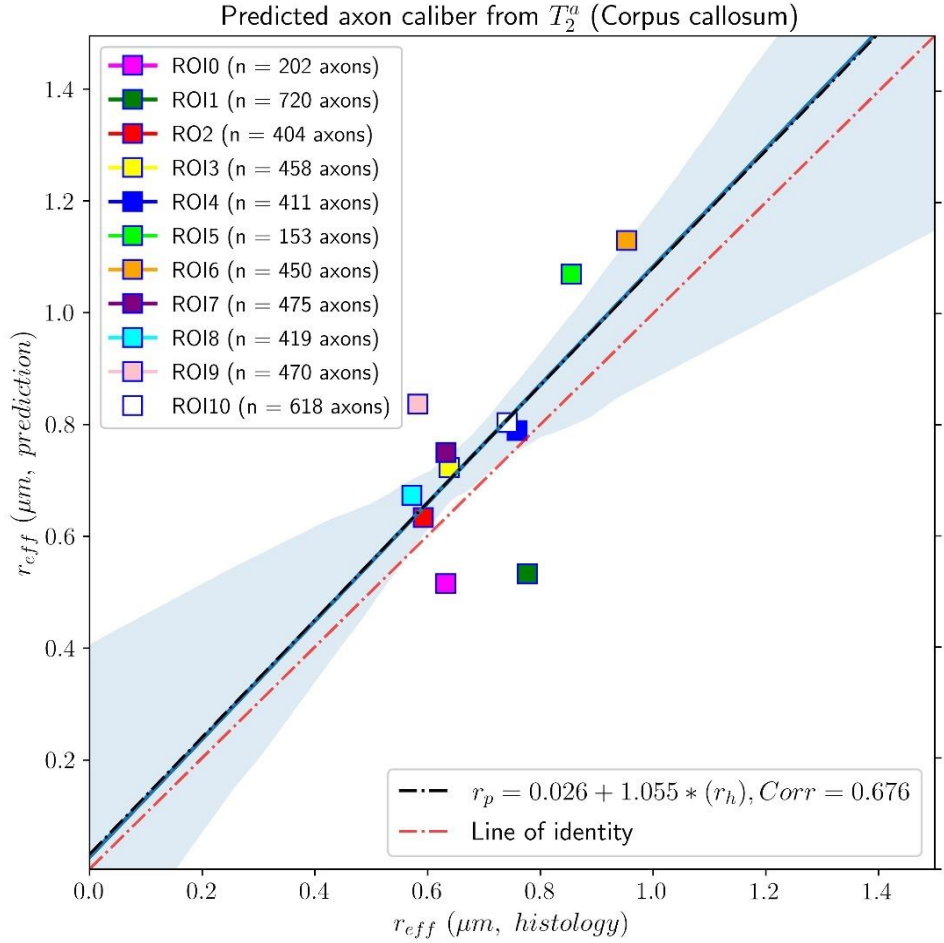

**Figure 6**. Linear fitting of the effective histological radius estimated from the second histological sample (Histology2) of the first histological dataset to the predicted radius from the intra-axonal $T_2$ times, calculated from the in vivo diffusion-$T_1$-$T_2$ MRI data. The scatter plot depicts the mean values computed for all the voxels inside the eleven corpus callosum (CC) regions, corresponding to ROI0-ROI10. The number of axons sampled for each CC ROI is displayed in the legend. The slope of the regression line was significantly different from zero (p=0.022).

Figures 7 and 8 show results from similar experiments using the intra-axonal $T_1$. Figure 7 depicts the linear fitting of the inverse of the mean intra-axonal $T_1$ per ROI estimated from in vivo diffusion-$T_1$-$T_2$ MRI data to the inverse of the mean effective radius corresponding to the Histology1 sample (Eq. (2)). The correlation coefficient between both variables was 0.755 lower than that previously found for the intra-axonal $T_2$ in Figure 5, and the p-value of the slope and the correlation did not reach statistical significance, p=0.25. From the estimated coefficients, we found the calibrated parameters to be $T_1^c \approx 870$ ms and $\rho_1 \approx 0.087$ nm/ms.

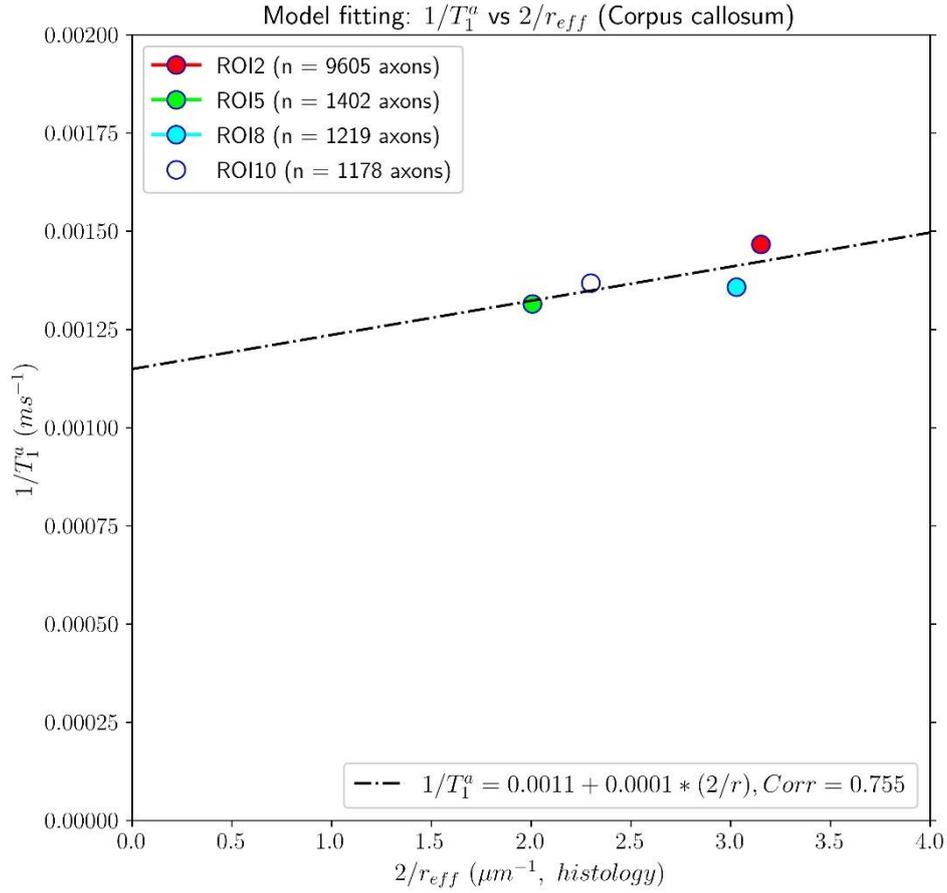

**Figure 7**. Linear fitting of the inverse of the intra-axonal $T_1$ times (y-axis) estimated from the in vivo diffusion-$T_1$-$T_2$ MRI data to the inverse of the inner axon radius (x-axis), measured from the first histological sample (Histology1) of the first histological dataset. The scatter plot depicts the mean values computed for all the voxels inside four corpus callosum (CC) regions, corresponding to ROI2, ROI5, ROI8, and ROI10. The number of axons sampled for each CC ROI is displayed in the legend. The intercept and slope of the regression line were 0.0011 and 0.000087. The p-value for the slope was not statistically significant (p=0.23).

The linear relationship between the effective mean axon radius estimated in the Histology2 sample and the radius predicted by using the intra-axonal $T_1$ times (Eq. (3)) is shown in Figure 8. The intercept and slope of the regression line were 0.064 μm and 1.002, respectively. The correlation coefficient was 0.628, and the slope was significant, p=0.039. When analyzing the subset of seven ROIs, excluding the four ROIs from the Histology1 sample, we obtained a new slope of 0.962 (p=0.16), which was not statistically significant. The intercept was 0.065 μm, and the correlation coefficient was 0.598.

Insert Figure 8 (1.5 columns)

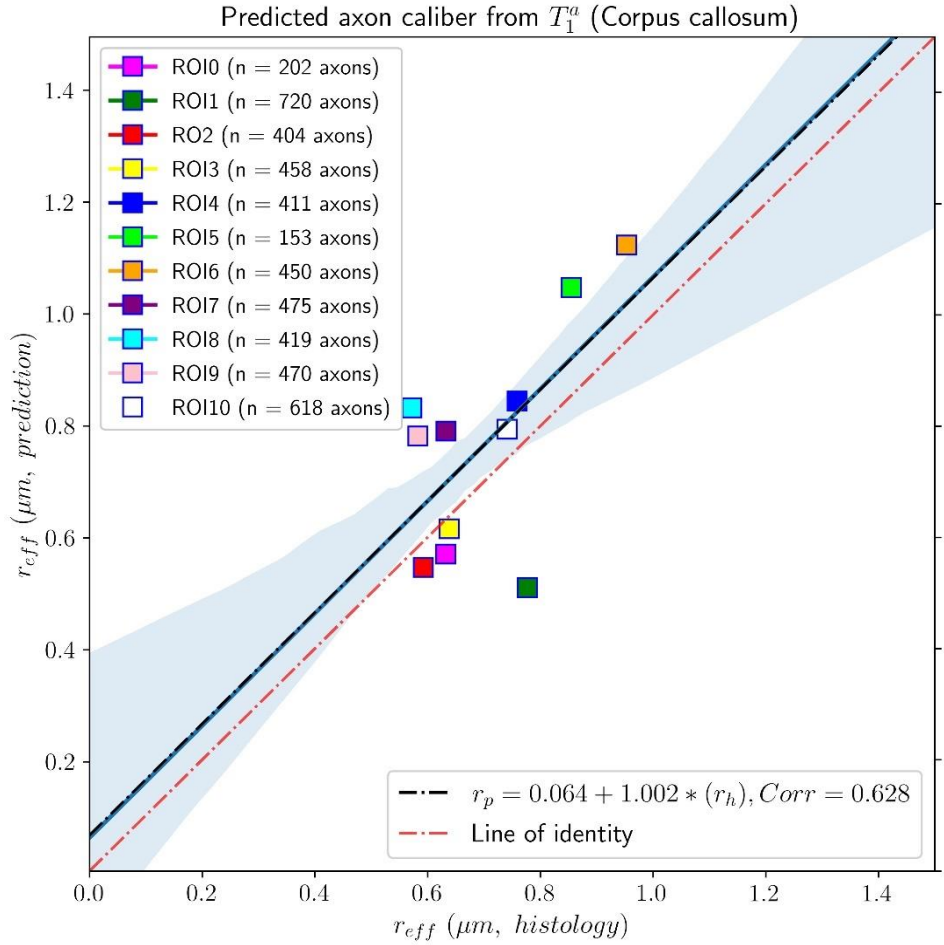

**Figure 8**. Linear fitting of the effective histological radius determined in the second histological sample (Histology2) of the first histological dataset to the predicted radius from the intra-axonal $T_1$ times estimated from the in vivo diffusion-$T_1$-$T_2$ MRI data. The scatter plot depicts the mean values computed for all the voxels inside the eleven corpus callosum (CC) regions, corresponding to ROI0-ROI10 in the Histology2 sample. The number of axons sampled for each CC ROI is displayed in the legend. The slope of the regression line was significantly different from zero (p=0.039).

Table 1 reports the mean histological effective axon radius per ROI and the predicted values from the intra-axonal $T_2$ and $T_1$ times, respectively. The predicted axon radius from both intra-axonal $T_2$ and $T_1$ times were very similar to each other. A linear fitting between both estimates revealed a slope close to one (0.947) and an intercept close to zero (0.041 µm). The slope was significantly non-zero (p=4e-5), and the correlation coefficient was 0.927.

Insert Table 1

**Table 1**. Mean effective radius (in $\mu m$) for each region of interest (ROI) in the Corpus Callosum. The anatomical location of each ROI is shown in Fig 3. The second row lists the radii corresponding to the Histology2 sample. The third and four rows report the predicted axon radii from the intra-axonal $T_2$ and $T_1$ times, respectively, estimated from the in vivo diffusion-$T_1$-$T_2$ MRI data.

| ROI | Histology | $T_2^a$ | $T_1^a$ |
|---|---|---|---|
| ROI0 | 0.632 | 0.515 | 0.571 |
| ROI1 | 0.777 | 0.533 | 0.511 |
| ROI2 | 0.592 | 0.634 | 0.547 |
| ROI3 | 0.638 | 0.723 | 0.616 |
| ROI4 | 0.759 | 0.789 | 0.844 |
| ROI5 | 0.855 | 1.069 | 1.047 |
| ROI6 | 0.953 | 1.129 | 1.123 |
| ROI7 | 0.633 | 0.750 | 0.791 |
| ROI8 | 0.572 | 0.673 | 0.832 |
| ROI9 | 0.583 | 0.836 | 0.782 |
| ROI10 | 0.741 | 0.803 | 0.794 |

**4.2 diffusion-$T_2$ and Histology1-Histology2-Histology3 data**

We complement the results presented in the previous section by reporting the predicted radii for the subjects scanned with the diffusion-$T_2$ MRI sequence and by including the Histology3 dataset. Notably, the parameters $T_2^c$ and $\rho_2$ were not recalibrated for these subjects; instead, we used the values estimated in the previous section.

The estimated intra-axonal $T_2$ values in the whole medial part of the CC for the three subjects were distributed in the following ranges 80-130 ms, 90-125 ms, and 85-115 ms, respectively.

In Figure 9, the predicted mean effective radius, derived from the intra-axonal $T_2$ times of the three subjects, is presented for all the CC ROIs. The figure also depicts the mean histological effective radius for the three histological samples (Histology1, Histology2, and Histology3).

To assess the validity of the calibrated parameters, which were estimated from the subject scanned with the diffusion-$T_1$-$T_2$ sequence, for the subjects scanned with the diffusion-$T_2$ sequence, we repeated the calibration process using the mean intra-axonal $T_2$ times estimated for the three subjects and the Histology1 sample as a reference, as before. The recalibrated parameters were remarkably similar to those obtained previously: $T_2^c \approx 127.17$ ms and $\rho_2 \approx 1.13$ nm/ms.

We compared the $T_2$-based predicted radii for the subject that underwent two scans, using both diffusion-relaxation MRI sequences, which values are reported in Table 1 and Figure 9 (as Subject 3). The linear fitting between both estimates produced a statistically significant regression line with a slope close to one (0.993, p<0.001) and an intercept close to zero (-0.029 µm). The correlation coefficient between both estimates was 0.884.

Insert Figure 9 (1.5 columns)

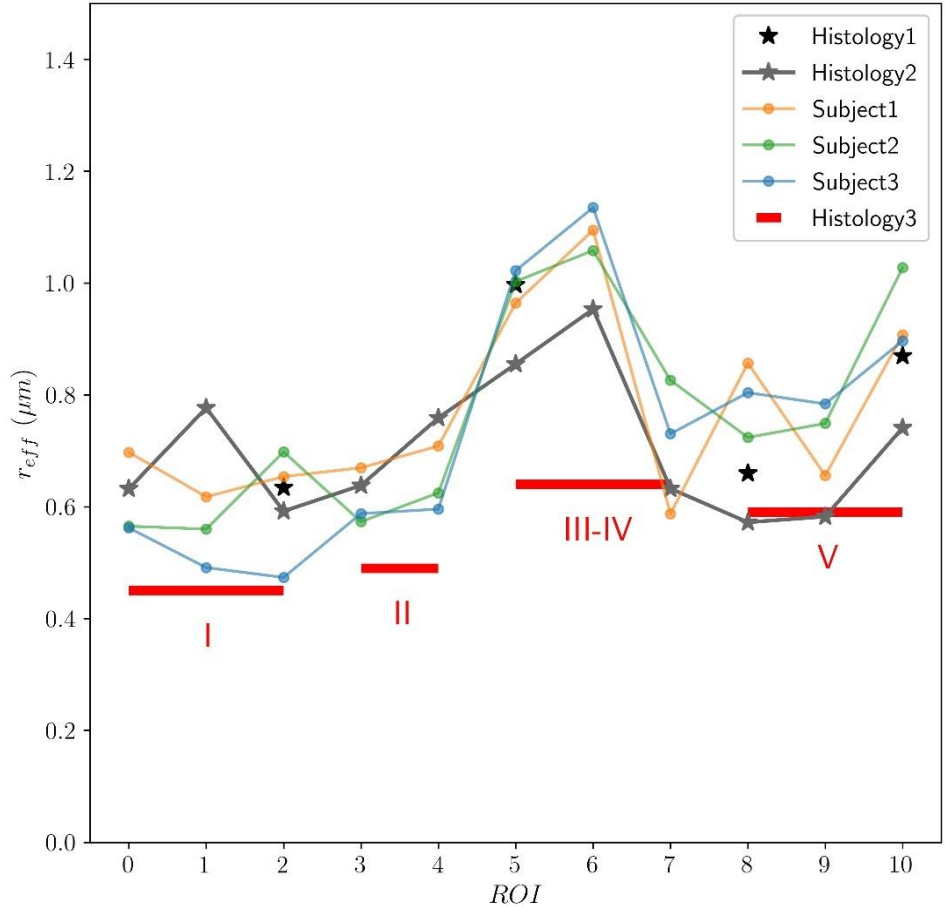

**Figure 9.** Predicted axon radius from intra-axonal $T_2$ times estimated from the in vivo diffusion-$T_2$ MRI data for the eleven ROIs (ROI0-ROI10) of the Histology2 sample. Additionally, as a reference, the mean effective histological radius calculated from the three histological samples (Histology1, Histology2, and Histology3) is also reported. Although the number and location of the ROIs used in the Histology3 sample differ from those employed in the Histology1-Histology2 samples, they can be regrouped to cover similar anatomical areas (see subsection "Histological samples" for more details). The histological and $T_2$-based radii follow the expected "low-high-low" trend in axon radii. The axon radii from the Histology1-Histology2 samples are consistently higher (about 25%) than those in the Histology3 sample.

Finally, we employed the calibrated model to predict the axon radius across the whole WM. Axial and sagittal slices of the voxel-wise $T_2$-based inner axon radius estimated for all the subjects are shown in Figure 10. The maps are approximately symmetrical, the spatial variability of the estimated radius is apparent in both slices, and all subjects show a similar pattern of small and big axons in the same anatomical regions.

Insert Figure 10 (1.5-2 columns)

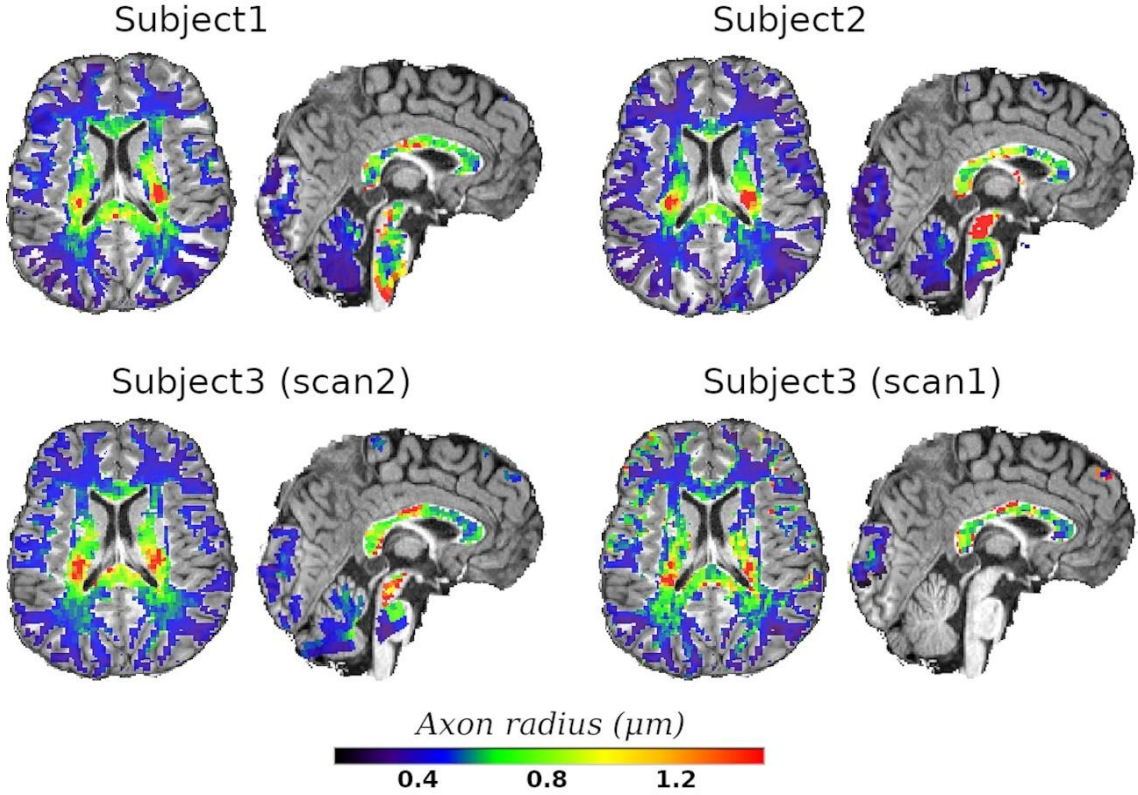

**Figure 10.** Axial and sagittal slices of the $T_2$-based inner axon radius for the three scanned subjects. Subject3 underwent two scans, with scan2 (46 slices) and scan1 (10 slices) representing the in vivo diffusion-T2 and diffusion-T1-T2 MRI data, respectively. All maps were normalised to the reference T1w image, where the histological CC ROIs were defined, and the predicted radii were plotted over the reference image. A white matter mask was used to suppress voxels in grey matter or cerebrospinal fluid.

## 5. Discussion

This proof-of-concept study shows that (1) the intra-axonal $T_2$ and $T_1$ relaxation times are highly modulated by the axon radius (see Figs. 5 and 7), as measured from histological data (see Fig. 3), (2) a simple surface-based relaxation model can explain this dependence (see Fig. 1), and (3) the intra-axonal relaxation times may also be sensitive to the smallest axons. Indeed, we did not observe a clear overestimation bias in the estimated axon radius (see Figs. 6, 8, and 9) in comparison to the histological values, as reported in previous dMRI studies (Assaf et al., 2008; Alexander et al., 2010; Dyrby et al., 2013; Horowitz et al., 2015) where only the largest radii might have been detected. This result suggests that our new approach may also be sensitive to differences in axon radius below the 'diffusion resolution limit' of ~1-2 μm. One possible explanation for this finding is that the intra-axonal $T_2$ times are not influenced by the strength of the diffusion gradients, as opposed to the intra-axonal radial diffusivities used to estimate axon radii in dMRI. Moreover, we found that the effective mean radius estimated by our approach, i.e., $r_{eff} \approx \langle r^2 \rangle / \langle r \rangle$, produces much smaller radii than those from diffusion models heavily weighted by the tail of the axon radius distribution, i.e., $r_{eff} \approx \left(\langle r^6 \rangle / \langle r^2 \rangle\right)^{\frac{1}{4}}$ (Burcaw et al., 2015; Veraart et al., 2020). This important result suggests that, from a modelling point of view, the employed diffusion-relaxation model may be more valuable than previous pure dMRI models for estimating axonal radii. The predicted mean effective radius obtained from the intra-axonal $T_2$ and $T_1$ times fell within a narrow range of 0.52-1.13 µm and 0.51-1.12 µm, respectively, which closely matched the range of histological axon radii (0.57-0.95 µm). The smallest predicted effective radii were observed in ROI1, ROI0, and ROI2, while the largest radii were found in ROI6 and ROI5, followed by ROI4, ROI9 and ROI10 (refer to Table 1). Nevertheless, we cannot rule out the possibility that the calibration step, informed by the histological values, may have reduced any potential overestimation effect.

Inspecting the estimated $T_2^a$ and $T_1^a$ relaxation maps (see Fig. 4), we notice that both relaxation times tend to be smaller in the genu and splenium of the CC than in the corticospinal tract (connecting the motor cortex to the spinal cord). Although these values could be affected by fibre orientation effects with respect to the B0 field (see the subsection "Orientation dependence on relaxation times" in the Appendix), the corticospinal tract is characterised by axons with larger inner radius (Aboitiz et al., 1992; Innocenti et al., 2014; Barakovic et al., 2021a). This observation agrees with multi-echo $T_2$ relaxometry studies showing that the intra- and extra-axonal $T_2$ times (and the myelin content) in the corticospinal tract are larger than in the CC, e.g., see (Yu et al., 2020; Canales-Rodríguez et al., 2021b, 2021c, 2021a). A consistent trend was observed in the $T_2$-based predicted axon radii for all three subjects, as shown in Figure 10. The voxel-wise maps in Figure 10 and the ROI-based estimates in Figure 9 agree with previous estimates derived from dMRI data acquired using much higher b-values (Veraart et al., 2021).

In agreement with our results, a previous multi-echo $T_2$ relaxometry study found a positive correlation between axon radius and $T_2$ (including both the intra- and extra-axonal compartments) in six samples of an excised and fixed rat spinal cord (Dula et al., 2010). Moreover, two previous experimental studies investigated the microstructural correlates of $T_1$ in white matter (Hofer et al., 2015; Harkins et al., 2016). In line with our findings, a significant correlation between $1/T_1$ and axon radius was reported by (Harkins et al., 2016) in white matter tracts of a rat spinal cord. Similarly, the analysis performed by (Hofer et al., 2015) found a tendency for the lowest $T_1$ in the genu of the human CC (composed of densely packed smaller axons) and the highest $T_1$ in the somatomotor region (dominated by fibres with large radii). In those studies, however, the estimated relaxation times characterise the relaxation process in the intra- and extra-axonal compartments combined. In contrast, we report a more specific relationship by analysing the intra-axonal relaxation times associated with the inner axon radius.

A multi-gradient-echo MRI model was proposed to estimate axon density based on the susceptibility-driven non-monotonic time-dependent MRI signal decay (Nunes et al., 2017). They employed a simple (phenomenological) general-linear model to predict the average axonal diameters using four modelled parameters, including the $T_2$ relaxation times of the intra- and extra-axonal compartments.

**5.1 Impact on previous and future studies**

Our study has important implications for previous and future dMRI studies of white matter microstructure. Previous studies, such as those by (Assaf et al., 2008; Alexander et al., 2010; Zhang et al., 2011; Dyrby et al., 2013; Xu et al., 2014; Daducci et al., 2015; Horowitz et al., 2015; Huang et al., 2015; Sepehrband et al., 2016b; Benjamini et al., 2016; Drobnjak et al., 2016; Romascano et al., 2020; Harkins et al., 2021; Herrera et al., 2022), estimated axon radius without considering any $T_2$ dependence, assuming the same $T_2$ for all axons and intra- and extra-axonal water compartments. This simplification may affect the estimation of the intra-axonal diffusivities from which the axon radii are derived. Alternatively, this issue could be attenuated by using sufficiently high *b*-values, as shown in studies by (Veraart et al., 2020; Pizzolato et al., 2022), which helps eliminate the contribution from the extra-axonal signal. However, this may be insufficient in voxels with a broad distribution of intra-axonal $T_2$ times. These multi-compartment models should be extended to include $T_2$ dependence, as discussed in studies by (Veraart et al., 2018; Lampinen et al., 2019; Tax et al., 2021). Recently, (L et al., 2022) demonstrated that more accurate estimates of neurite size could be obtained by investigating the coupling between relaxation rate and diffusivity using multi-TE diffusion-relaxation MRI data. For further discussion on this issue, please refer to the Appendix subsection "Is the intra-axonal relaxation process mono-exponential and time-independent?".

**5.2 Underlying assumptions and confounding factors**

The proposed diffusion-relaxation model specifically applies to WM regions composed of myelinated axons, where the exchange of water molecules and other macromolecules and elements (such as iron/ferritin) between the intra- and extra-axonal spaces is negligible. In the human brain's CC, for example, more than 95% of axons in most regions are myelinated (Aboitiz et al., 1992). While the non-myelinated portions of the axon (i.e., nodes of Ranvier) contain a high density of voltage-gated ion channels that facilitate ion passage across the axonal membrane, including K+ and Na+, which is associated with a concomitant water flux (Badaut et al., 2002), the myelinated portions of the axon (i.e., internodes) are not exposed to the extracellular environment. Although the axonal membrane in the nodes of Ranvier is semipermeable to small diffusing molecules, such as water, the internodes' length is significantly greater (∼100 times the outer axon diameter (Hursh, 1939; Rushton, 1951)) than the nodes of Ranvier (∼1 μm (Arancibia-Cárcamo et al., 2017)). As a result, most multi-compartment $T_2$ (Lancaster et al., 2003; Deoni et al., 2013) and 'standard' dMRI models (see (Novikov et al., 2019) for a review) assume that the measured MRI signal is not significantly affected by the inter-compartmental molecular exchange in WM regions composed of myelinated axons.

Therefore, it is important to note that our model is unsuitable for GM or WM regions affected by demyelination processes, such as in Multiple Sclerosis, or any pathological condition with an increase in intra-axonal iron. These conditions can significantly reduce intra-axonal relaxation times and the estimated radii, rendering our model invalid. However, it is worth noting that we use long TEs in our model. If the intra-axonal $T_2$ time of a given axon is significantly reduced (e.g., below 20-40ms) due to external factors, the contribution of this axon to the overall voxel-wise measured signal will be greatly diminished.

However, it is interesting that our calibration approach could also be extended to cases where water exchange between intra- and extra-axonal spaces is non-negligible, provided the exchange is similar across axons with different radii. In such cases, the effect of the exchange on the observed intra-axonal relaxation times can be modelled by a global scaling of the cytoplasmic relaxation time, which is accounted for during calibration.

A more suitable approach for modelling systems that are coupled by means of a relaxation exchange process could be based on the Bloch-McConnell equations (McConnell, 1958), which generalise the relaxation model employed in this study (Eqs. (5) and (6)). However, fitting such a model requires estimating additional parameters, including membrane permeability and extra-axonal relaxation times, which may be prone to numerical degeneracies. Additionally, the MRI acquisition time required for fitting the Bloch-McConnell model (using both high and low *b*-values) is longer than that required for our proposed model.

A study on human postmortem brains revealed that $T_2^*$ is more sensitive than $T_2$ to changes in WM iron concentration (Langkammer et al., 2010). While it is established that the macromolecular and iron content is altered in certain pathologies (Stüber et al., 2014), more research is required to understand how these abnormalities affect the intra-axonal space and how they can impact the intra-axonal relaxation times.

We assume that signals measured at very high *b*-values are primarily attributable to the intra-axonal space, given that the signals from free-water and extra-axonal compartments decay more rapidly with the *b*-value (Veraart et al., 2020). To further suppress signals from tissue compartments with very short $T_2$s, such as myelin water (Mackay et al., 1994) and other confined water molecules, we also used long TEs. Hence, the resulting signals are expected to come from intra-axonal water molecules. However, there are other 1D-stick-like structures in the WM, such as the radiating processes of astrocytes, which can have large diameters that might contaminate the resulting signals (Veraart et al., 2020, 2021), as well as cell nuclei, vacuoles, and other restricted compartments (Andersson et al., 2020). Therefore, further studies are necessary to understand the potential effects of these compartments on the measured $T_2$ and predicted radii.

### 5.3 Acquisition sequences

The diffusion-$T_1$-$T_2$ sequence was implemented to investigate the impact of axon radius on the intra-axonal $T_1$ and $T_2$ times independently. Our results demonstrate that both relaxation times are sensitive to changes in axon radius, with $T_2$ exhibiting a slightly higher sensitivity. Consequently, we can obtain two separate estimates of axon radius using the relaxation times calculated from this sequence (see Table 1). However, this is not our recommended acquisition protocol due to the long acquisition time required. Alternatively, a more practical approach for estimating axon radius is to use the diffusion-$T_2$ sequence. A faster version of this sequence could be implemented by utilising only two TEs, although the resulting estimates may be more affected by underlying noise. This possibility shall be investigated in future studies.

When implementing these sequences, it is important to identify the optimal *b*-value to attenuate the extra-axonal signal. Based on in vivo human brain data and numerical experiments using analytical equations, the general rule of thumb is that a *b*-value in the range of 4000-6000 s/mm$^2$ is sufficient (Jensen et al., 2016; McKinnon and Jensen, 2019). In our study, we used the highest *b*-value within this range. However, it is worth noting that determining the optimal *b*-value involves a trade-off influenced by the SNR, which is affected by other sequence parameters, including the TE and voxel size. Our data were acquired using the Connectom 3T scanner at CUBRIC, which has been previously used to collect data with *b*-values up to 30000 s/mm$^2$ (Veraart et al., 2020, 2021). Ultra-high *b*-values with very strong diffusion gradients are necessary for pure dMRI models to improve sensitivity to smaller axon radii (Nilsson et al., 2017). However, our sequences do not require *b*-values larger than 6000 s/mm$^2$ because all the necessary information is derived from the relaxation times, which depend on the TEs/TIs.

### 5.4 Main limitations and future studies

While our study provides valuable insights into the relationship between axon radii and MRI relaxation times, it is important to acknowledge some limitations. First, the in vivo diffusion-relaxation MRI data and postmortem histological samples were obtained from different subjects of different ages and genders. Although some studies suggest that there are no sex differences in the fibres composition of the corpus callosum (Aboitiz et al., 1992), others have found age-related changes in axon size (Aboitiz et al., 1996), which may affect the comparison between the postmortem and in vivo measurements. Therefore, the estimated relaxation times of the cytoplasmic water and surface relaxivities must be considered as approximated guide values.

Second, the histological analysis of the second and third histological samples (Histology2 and Histology3, covering eleven and five CC sectors, respectively) are based on a reduced number of axons compared to the first (Histology1) sample. This may introduce sampling biases that could affect the accuracy of the histological radius estimates. An extended discussion is provided in the Appendix subsection "Histological tissue shrinkage and sampling issues". As such, a perfect agreement between the effective histological radius and the predicted MRI-based radius was not expected.

Third, the analysis was confined to the mid-sagittal plane of the CC. Therefore, the estimated mean cytoplasmic relaxation times and surface relaxivities are specific to this region. It is possible that different values may be obtained if other white matter tracts were included in the analysis. However, the extension of the analysis to other regions would require modelling the orientation susceptibility effects, which is beyond the scope of this proof-of-concept study. For more details, refer to the Appendix subsection "Orientation dependence on relaxation times".

Fourth, our study had a relatively small number of data points available for computing correlations, with only four ROIs to implement the calibration. This limited sample size restricts the statistical power and precision of the estimated correlations, leading to increased uncertainty and decreased reliability of the findings. While it is generally recommended to account for multiple comparisons to reduce the risk of false-positive findings, we opted not to implement such correction. Given the exploratory nature of our study, we prioritised sensitivity over stringent control of false positives. Consequently, our findings should be interpreted cautiously, requiring further validation in independent studies. However, for completeness, we report that if we correct our results for multiple comparisons using the Bonferroni method, only two analyses survive the correction: the correlation of the radii estimated using the $T_2$ and $T_1$ relaxation times reported in Table 1 and the $T_2$-based predicted radii for the subject who underwent two scans, using the two diffusion-relaxation MRI sequences employed in this study.

Fifth, the proposed model is not suitable for GM or WM regions affected by demyelination processes or any pathological condition increasing intra-axonal iron. These conditions can significantly reduce intra-axonal relaxation times and the estimated radii, rendering the model invalid. A more detailed discussion of these underlying assumptions and confounding factors can be found in the previous subsection, "Underlying assumptions and confounding factors", in the Discussion section.

Sixth, the estimation of intra-axonal $T_2$ from the spherical mean of the strongly diffusion-weighted signal may be subject to bias due to the presence of isotropically-restricted compartments, including cell nuclei and vacuoles (Andersson et al., 2020). However, this issue can be mitigated by utilizing the spherical variance instead (Pizzolato et al., 2022). For more detailed information, please refer to the Appendix subsection "The effect of spherical cells: spherical mean vs spherical variance."

Seventh, although the spherical mean signal is not affected by the presence of fibre crossings and orientation dispersion (Lindblom et al., 1977; Callaghan et al., 1979; Kaden et al., 2016a), it is influenced by the orientation susceptibility effect. In other words, the measured signal still depends on the angle between the B0 magnetic field and the fibre orientation. In our study, the regions of interest were located in the medial part of the CC, where the angle between the B0 vector field and the nerve fibres remains relatively constant. More details on this topic can be found in the Appendix subsection "Orientation dependence on relaxation times.

To better assess the generalisability of our approach, further validation studies are necessary. In particular, we plan to test our method using biomimetic phantoms with known ground truth (Hubbard et al., 2015; McHugh et al., 2018; Huang et al., 2021; Zhou et al., 2021) and ex-vivo data from the same brains and multiple white matter regions. Such datasets would allow us to investigate whether the cytoplasmic relaxation times are truly independent of axon radius (see the Appendix subsection "Is the cytoplasmic T2 constant?"). This could be achieved by repeating the calibration process using different subsets of ROIs and comparing the resulting estimates. However, obtaining sufficient histological ROIs and measured axons per ROI will be crucial to minimize sampling bias and get robust results not affected by noise. Additionally, including data from the same brains (e.g., from non-human studies) will enable us to guarantee that we are studying the same axonal bundles.

An interesting future direction would be to utilise bundle-specific intra-axonal $T_2$ values (Barakovic et al., 2021b) to estimate bundle-specific inner axon radius, which could potentially resolve multiple axonal radii per voxel. This approach may potentially predict axon radius beyond the current dMRI resolution limit using clinical scanners. However, one limitation of translating the diffusion-relaxation MRI sequence to clinical scanners is the decreased signal-to-noise ratio resulting from using high *b*-values and long TEs. One potential solution to mitigate this could be to reduce the *b*-value to 4000-5000 s/mm$^2$ and use numerical simulations to determine the optimal range of TEs, based on the intra-axonal relaxation times reported in this study and the expected noise range.

Despite these limitations, our study provides a promising approach for estimating axon radii and understanding their relationship with MRI relaxation times. Future studies could address these limitations and expand the analysis to other brain regions to further validate the technique.

## 6. Code and data availability statement

The datasets used in this study and the Python code can be made available upon request from the corresponding authors, subject to the following terms and conditions. The mean effective histological radii of the Histology1, Histology2, and Histology3 samples are reported in Table 1 and Figure 9, respectively. We can also share any other derived metric from the Histology1-Histology2 samples. Additional results for the Histology2 and Histology3 samples are available in (Caminiti et al., 2009) and (Wegiel et al., 2018), respectively. The MRI data will be available upon signing a data-sharing agreement with Cardiff University. Finally, we can provide the Python scripts used in this study upon request.

## 7. Acknowledgements

EJC-R is supported by the Swiss National Science Foundation (SNSF), Ambizione grant PZ00P2_185814. DKJ is supported by a Wellcome Trust Investigator Award (096646/Z/11/Z) and a Wellcome Trust Strategic Award (104943/Z/14/Z). TBD is supported by the European Research Council (ERC), Grant/Award Number: 101044180. CMWT is supported by the Wellcome Trust (215944/Z/19/Z) and the Dutch Research Council (NWO, 17331). For the purpose of open access, the author has applied a CC-BY public copyright licence to any author accepted manuscript version arising from this submission. The presented study is a tribute to Professor Giorgio Innocenti (1946-2021).

## 8. Appendix: effective axon radius

We derive the mean effective radius that can be estimated from the intra-axonal $T_2$ and $T_1$ relaxation times. For simplicity, we will separately analyse the components of the measured signals that exclusively depend on $T_2$ and $T_1$.

**8.1 Axon radius estimated from $T_2^a$**

The signal arising from the intra-axonal compartments is the sum of signals from the spins inside all axons. The measured T2-weighted signal for a given echo time *TE* is

$$M(TE) = k \sum_{i=1}^{P} N_i \exp\left(-\frac{TE}{T_2^i}\right), \qquad (7)$$

where *P* is the total number of axons, $N_i$ is the number of spins inside the $i^{th}$ axon with transverse relaxation time $T_2^i$, and *k* is a constant that depends on the sequence/scanner.

Assuming that the proton density (*PD*) does not depend on the axon radius, then

$$PD = \frac{N_i}{\pi r_i^2 h} = \frac{\sum_{i=1}^{P} N_i}{\sum_{i=1}^{P} \pi r_i^2 h} = \frac{N_t}{\sum_{i=1}^{P} \pi r_i^2 h}. \qquad (8)$$

where $\pi r_i^2 h$ is the volume occupied by the $i^{th}$ axon, *h* is the axon length, and $N_t$ is the total number of spins in the intra-axonal space.

From Eq. (8) we obtain the following simplified relationship:

$$N_i = N_t \frac{r_i^2}{\sum_{i=1}^{P} r_i^2}, \qquad (9)$$

By plugging Eq. (9) into Eq. (7) we get

$$M(TE) = kN_t \sum_{i=1}^{P} \left(\frac{r_i^2}{\sum_{i=1}^{P} r_i^2}\right) \exp\left(-\frac{TE}{T_2^i}\right). \qquad (10)$$

We estimate a single intra-axonal T2 per voxel, which is equivalent to assuming that all the T2s in Eq. (10) are equal to $T_2^a$ (i.e., and hence that all axons in the voxel have the same radius $\bar{r}$); in that case, Eq. (10) becomes

$$M(TE) \approx kN_t \sum_{i=1}^{P} \left(\frac{1}{P}\right) \exp\left(-\frac{TE}{T_2^a}\right) = kN_t \exp\left(-\frac{TE}{T_2^a}\right). \qquad (11)$$

To understand how the distribution of axon radii in Eq. (10) affects the apparent $T_2^a$ in Eq. (11), we use the following approximation

$$kN_t \exp\left(-\frac{TE}{T_2^a}\right) \approx kN_t \sum_{i=1}^{P} \left(\frac{r_i^2}{\sum_{i=1}^{P} r_i^2}\right) \exp\left(-\frac{TE}{T_2^i}\right). \tag{12}$$

After plugging the surface-based relaxation model in Eq. (1) and removing common terms on both sides of the previous equation, we get

$$\exp\left(-\frac{2TE\rho_2}{\bar{r}}\right) \approx \sum_{i=1}^{P} \left(\frac{r_i^2}{\sum_{i=1}^{P} r_i^2}\right) \exp\left(-\frac{2TE\rho_2}{r_i}\right), \tag{13}$$

where we cancelled the contribution from the cytoplasmic $T_2^c$, which appears on both sides of the equation.

The exponential terms $2TE\rho_2/r_i$ are small (according to our data and results <0.5), so we can expand the exponentials in Taylor series using a first-order approximation as

$$\begin{aligned}
1 - \frac{2TE\rho_2}{\bar{r}} &\approx \sum_{i=1}^{P} \left(\frac{r_i^2}{\sum_{i=1}^{P} r_i^2}\right)\left(1 - \frac{2TE\rho_2}{r_i}\right) \\
&= \sum_{i=1}^{P} \left(\frac{r_i^2}{\sum_{i=1}^{P} r_i^2}\right) - \sum_{i=1}^{P} \left(\frac{r_i 2TE\rho_2}{\sum_{i=1}^{P} r_i^2}\right) \\
&= 1 - 2TE\rho_2 \sum_{i=1}^{P} \left(\frac{r_i}{\sum_{i=1}^{P} r_i^2}\right).
\end{aligned} \tag{14}$$

Therefore,

$$\begin{aligned}
\frac{1}{\bar{r}} &\approx \sum_{i=1}^{P} \left(\frac{r_i}{\sum_{i=1}^{P} r_i^2}\right), \\
\bar{r} &\approx \frac{\sum_{i=1}^{P} r_i^2}{\sum_{i=1}^{P} r_i} = \frac{\langle r^2 \rangle}{\langle r \rangle},
\end{aligned} \tag{15}$$

This is the expression that we used to correct the histological radii, which is the mean effective radius estimated from this relaxation model.

## 8.2 Axon radius estimated from $T_1^a$

Following a similar approach, the measured $T_1$-weighted signal for a given $TI$ is

$$M(TI) = k\sum_{i=1}^{P} N_i \left|1 - 2\exp\left(-\frac{TI}{T_1^i}\right)\right| \\ = kN_t \sum_{i=1}^{P} \left(\frac{r_i^2}{\sum_{i=1}^{P} r_i^2}\right)\left|1 - 2\exp\left(-\frac{TI}{T_1^i}\right)\right|. \tag{16}$$

Note that we neglected the *TR* dependence because, in practice, this experimental parameter is much higher than the intra-axonal $T_1$, and its contribution is minor.

As we estimate a single apparent intra-axonal $T_1$ per voxel, our model is equivalent to assuming that all the $T_1$s are equal to $T_1^a$ (i.e., all axons in the voxel have the same radius $\bar{r}$); thus, Eq. (16) becomes

$$M(TI) = kN_t \left|1 - 2\exp\left(-\frac{TI}{\overline{T_1}}\right)\right|. \tag{17}$$

To investigate how the distribution of axon radii in Eq. (16) affects the apparent $T_1^a$ in Eq. (17), we use the approximation

$$kN_t \left|1 - 2\exp\left(-\frac{TI}{\overline{T_1}}\right)\right| \approx kN_t \sum_{i=1}^{P} \left(\frac{r_i^2}{\sum_{i=1}^{P} r_i^2}\right)\left|1 - 2\exp\left(-\frac{TI}{T_1^i}\right)\right|. \tag{18}$$

After plugging the surface-based relaxation model in Eq. (2) and removing common terms on both sides of Eq. (18) we obtain

$$\exp\left(-\frac{2TI\rho_1}{\bar{r}}\right) \approx \sum_{i=1}^{P} \left(\frac{r_i^2}{\sum_{i=1}^{P} r_i^2}\right)\exp\left(-\frac{2TI\rho_1}{r_i}\right), \tag{19}$$

Note that Eq. (19) is similar to Eq. (13). Hence, we can get the same relationship given by Eq. (15) after using the first-order Taylor series approximation, which is justified by the small values of the exponential terms $2TI\rho_1/r_i$ (according to our MRI acquisition parameters and results <0.3).

**8.3 Histological tissue shrinkage and sampling issues**

The histological datasets were inspected to investigate the trend in axon radii. As expected, the data followed the "low-high-low" pattern in axon radii, as shown in Figure 9. However, the mean effective histological radii differed between the samples. The axon radii from the Histology1-Histology2 samples were about 25% higher than those in the Histology3 sample. These differences could be due to genuine anatomical variations between the postmortem brains or related to the histological procedures and corresponding tissue shrinkage factors. The $T_2$-based predicted radii in all subjects followed a similar "low-high-low" pattern closer to the values measured in the Histology1 sample, as this was the calibration sample.

In this study, the histological samples were not corrected for tissue shrinkage, which can affect the accuracy of the estimated axon radii. Consequently, the in vivo axons may be thicker than the reported histological values (Barazany et al., 2009; Horowitz et al., 2015). The extent of tissue shrinkage can vary widely depending on the used histological preparation techniques, with reported shrinkage factors ranging from 1-65% (Lamantia and Rakic, 1990; Aboitiz et al., 1992; Houzel et al., 1994; Riise and Pakkenberg, 2011). It is also unclear if shrinkage affects all brain axons equally, as previous research has shown varying shrinkage levels in different cellular compartments (Hursh, 1939). However, there is currently limited knowledge about the effects of shrinkage on CC axons in the human brain (Innocenti et al., 2019). Please refer to (Dyrby et al., 2018) for further information on tissue shrinkage issues.

Sampling biases can impact histological radii measurements. One issue is that only a small amount of tissue is typically sampled, so the microstructure properties of these regions may not accurately represent properties in other regions within the ROIs (Assaf et al., 2008). Another issue is that larger axons can influence the mean effective radius more than the mean radius of the distribution. Since larger axons are less common, accurately detecting their proportions in a sample requires measuring a larger number of axons. We observed this effect in the Histology1 and Histology2 samples, where the effective radii in the four ROIs used in Histology1 (which had denser spatial sampling) were consistently higher than those in the same ROIs measured in the Histology2 sample (see Figure 9).

For these reasons, the presented histological results should not be considered the definitive "ground truth". Future studies should aim to identify optimal histological procedures, such as those suggested by (Sepehrband et al., 2016b), and also explore the use of neural network approaches for automatic measurement of tens of thousands of axons per ROI to reduce sampling biases (Mordhorst et al., 2022). It is also worth noting that because the proposed calibration approach uses histological data as a reference, the predicted radii are relative to the specific histological sample employed.

**8.4 The effect of spherical cells: spherical mean vs spherical variance**

A recent study showed that isotropically-restricted compartments might bias the intra-axonal $T_2$ estimated from the spherical mean of the strongly diffusion-weighted signal (Pizzolato et al., 2022). Thus, our estimates could be partially affected by cell nuclei, vacuoles, and other types of structures in the white matter (Andersson et al., 2020). As a remedy for that problem, it was proposed to use the spherical variance (Pizzolato et al., 2022) as a 'filter' since the spherical variance of an ordered axon bundle would be high, but in an isotropic component would be close to zero. Although the results obtained in that study are promising, the spherical variance is more sensitive to noise than the spherical mean. Moreover, a larger number of diffusion gradient directions than that used in our study is necessary to employ this novel technique (48 vs >96 in (Pizzolato et al., 2022)). In future studies, we plan to acquire dMRI data using a higher angular resolution to compare both techniques' outputs and filter out any contribution from spherical cells.

**8.5 Is the cytoplasmic T2 constant?**

The cytoplasmic $T_2$ may be influenced by the intra-axonal microstructure, such as the number of organelles and the density of cytoskeletal elements such as neurofilaments, microtubules, and actin, as well as the chemical composition, including the type and density of macromolecules and water content.

Numerous morphometric studies have provided evidence of a linear correlation between neurofilament and microtubule numbers and axonal cross-sectional area (Friede and Samorajski, 1970; Hoffman et al., 1987). These studies suggest that myelinated axons contain more neurofilaments than microtubules and that the axon radius adjusts to maintain a constant density of neurofilaments. It was demonstrated that this relationship is regulated by the relative degree of phosphorylation of the mid-sized and heavy neurofilaments (Rao et al., 1998). Furthermore, the myelin-associated glycoprotein is implicated in the signalling cascade controlling neurofilament phosphorylation (Lunn et al., 2002) and axon radius. As neurofilaments are the more abundant cytoskeletal elements and their density is nearly constant in myelinated axons with different radii, we do not anticipate a relationship between cytoplasmic $T_2$ and axon radius mediated by neurofilament density in the axonal cytoskeleton.

However, a previous study using electron probe x-ray microanalysis (LoPachin et al., 1991) measured the concentrations of biologically relevant elements (such as Na, P, S, Ca, CI, K, and Mg, in mmol/kg dry or wet weight) and water content in the axoplasm of rat optic nerve myelinated axons. The study found that dry and wet weight concentrations of Na, P, S and Ca were not dependent on the axonal radius. In contrast, the axoplasmic concentration of K, CI, and Mg was related to axon radius. Furthermore, the water content in medium and large axons was similar (between 91% and 92%) but slightly reduced in small axons (89%). These findings suggest that the chemical composition of the axoplasm depends on the axon radius (LoPachin et al., 1991). Therefore, until the effect of K, CI, and Mg on intra-axonal relaxation times is clarified, the surface-based relaxation model employed (i.e., Eqs. (1) and (2)) should be regarded as a first-order approximation.

Despite this limitation, our findings (refer to Figures 5 and 7) suggest a linear relationship between the inverse of intra-axonal relaxation times and axon radius, consistent with predictions made by the surface-based relaxation model we employed. Our empirical results demonstrate that the calibration step enables us to estimate the mean axon radius in various regions of the midsagittal CC (refer to Figures 6, 8, and 9). As we did not observe any significant nonlinear relationships between intra-axonal relaxation times and axon radii (within the range of measured radii in the midsagittal CC), we conclude that any nonlinear dependence is weak and can be disregarded. Hence, either the cytoplasmic relaxation times remain constant, as assumed in this study, or they vary linearly with axon radius. In the following, we present some examples where the calibrated model could predict the axon radius accurately, even if the cytoplasmic relaxation times depend on the axon radius.

Let us consider two distinct scenarios: In the first case, the cytoplasmic $T_2$ increases with the radius (similar to the observed trend for intra-axonal $T_2$ time), while in the second case, it decreases. The former corresponds to the model

$$\frac{1}{T_2^c(r)} = \frac{1}{T_2^{const}} + \frac{k}{r}, \quad (20)$$

where $T_2^{const}$ is a constant term common to all axons and $k$ is a constant quantifying how fast the cytoplasmic $T_2$ changes with $r$. By plugging this equation into our relaxation model (Eq. (1)), we get a similar model with redefined parameters

$$\frac{1}{T_2^a} = \frac{1}{T_2^{const}} + \frac{\gamma}{r}, \quad (21)$$

where $\gamma = 2\rho_2 + k$. Althout the new parameters $T_2^{const}$ and $\gamma$ cannot be interpreted as the cytoplasmic $T_2$ and surface relaxivity, they can be estimated by employing our calibration approach. Therefore, the resulting model would be equally valid for predicting axon radius.

The second case corresponds to a model that predicts a decrease in $T_2^c(r)$ for larger axons

$$\frac{1}{T_2^c(r)} = \frac{1}{T_2^{const}} + kr. \tag{22}$$

After plugging Eq. (22) into Eq. (1) and regrouping terms, the relaxation model becomes

$$\frac{1}{T_2^a} = \frac{1}{T_2^{const}} + \frac{(kr^2 + 2\rho_2)}{r}. \tag{23}$$

In our experiments, we observed a net reduction of $T_2^a$ with $r$. Hence, the surface relaxivity term must dominate the relaxation over $k$, i.e., $2\rho_2/r > kr$ for the range of measured radii. The modified parameters to be calibrated in this model are $T_2^{const}$ and $\gamma = 2\rho_2 + kr^2$. In this case, our model only provides a good approximation if the previous inequality becomes $2\rho_2/r \gg kr$.

It is important to note that the models presented in this section (Eqs. (20)-(23)) are hypothetical and were discussed to illustrate the flexibility and limitations of the calibration approach in cases where the underlying assumptions are not met. Similar results can be obtained by using the intra-axonal $T_1$ time or assuming a surface relaxivity that depends on the radius.

**8.6 Orientation dependence on relaxation times**

By computing the spherical mean of the diffusion signal, the resulting orientation-averaged signal is independent of the fibre orientation distribution and thus is not affected by the presence of fibre crossings and varying levels of fibre orientation dispersion (Lindblom et al., 1977; Callaghan et al., 1979; Kaden et al., 2016a). However, the spherical mean does not eliminate the dependence on the orientation susceptibility effects, i.e., the measured signal still depends on the angle between the B0 magnetic field and the fibre orientation. Some previous studies have reported this orientation dependence for both the $T_2^*$ and $T_2$ (Oh et al., 2013; Aggarwal et al., 2016; Gil et al., 2016) and $T_1$ (Knight and Kauppinen, 2016; Knight et al., 2017, 2018; Schyboll et al., 2018, 2020). Notably, while (McKinnon and Jensen, 2019) reported a significant intra-axonal $T_2$ orientation effect, a recent study found that extra-axonal $T_2$ is more affected by this phenomenon than intra-axonal $T_2$ (Tax et al., 2021). Given these inconsistent findings, further research is needed to determine whether the orientation-dependent $T_2$ effect is significant enough to be considered in this model.

In our study, the regions of interest were located in the medial part of the CC, where the angle between the B0 vector field and the nerve fibres remains relatively constant. Therefore, our findings are not likely affected by B0-orientation-related bias. However, the orientation effect should be modelled in brain regions with different fibre orientations, as it may affect the estimation. Despite this potential limitation, in Figure 10, we present $T_2$-based radius images across the entire white matter, showing the spatial variability of estimated radii across different regions, especially in the sagittal slices depicting the midsagittal CC cross-sections. The estimates from all subjects demonstrate a similar concordant pattern, as well as the maps of the same subject (Subject3) obtained from the two diffusion-relaxation MRI sequences, although some differences are noticeable due to the different voxel sizes used in both acquisitions.

**8.7 Is the intra-axonal relaxation process mono-exponential and time-independent?**

A recent theoretical formulation by (Kiselev and Novikov, 2018) demonstrated how the interplay between diffusion and spin dephasing in a heterogeneous environment could produce a non-mono-exponential time-dependent transverse relaxation signal. While this effect may be significant for short TEs, our relatively long TEs (i.e., >73 ms) and diffusion times ($\Delta$=22 ms, $\delta$=8 ms) used in this study (compared to the small intra-axonal space where the restricted diffusion process takes place) indicate that a mono-exponential signal relaxation is expected for the spins inside each axon.

In our study, we estimated a single intra-axonal relaxation time per voxel. However, if axon radii are distributed with non-negligible variance, a more complete formulation must consider distributions of relaxation times. Estimating a nonparametric distribution of relaxation times is problematic from a practical point of view because a large number of TEs/TIs would be required. Nevertheless, an approach similar to that introduced in AxCaliber (Assaf et al., 2008) could be adopted by assuming a parametric form for such distributions, as shown in (Sepehrband et al., 2016a). Future studies should investigate this generalization further.